\documentclass[cits]{PoS}

\usepackage{color}
\usepackage{upgreek}
\usepackage{amsmath}

\usepackage{url}

\title{Localizations of Fast Radio Bursts on milliarcsecond scales}

\ShortTitle{Localizations of Fast Radio Bursts on milliarcsecond scales}

\author{\speaker{B.~Marcote} and Z.~Paragi\\
	Joint Institute for VLBI ERIC, Oude Hoogeveensedijk 4, 7991 PD Dwingeloo, The Netherlands\\
        E-mail: \email{marcote@jive.eu}
}

\abstract{Fast Radio Bursts (FRBs) are transient sources that emit a single radio pulse with a duration of only a few milliseconds. Since the discovery of the first FRB in 2007, tens of similar events have been detected. However, their physical origin remains unclear, and a number of scenarios even larger than the number of known FRBs has been proposed during these years.
The detection of multiple bursts in FRB~121102 excluded all cataclysmic scenarios, at least for this particular FRB. The presence of these repeating bursts allowed us to perform a precise localization of the source with the Karl G.\ Jansky Very Large Array (VLA) and the European VLBI Network (EVN). Optical observations with Keck, Gemini and {\em HST} unveiled the host to be a low-metallicity star-forming dwarf galaxy located at a redshift of 0.193. The EVN results showed that the bursts are co-located (within a projected separation of $< 40$ pc) to a compact and persistent radio source with a size of $< 0.7$ pc and located within a star-forming region.
This environment resembles the ones where superluminous supernovae or long-duration gamma-ray bursts are produced. Although the nature of this persistent source and the origin of the bursts remain unknown, the scenarios considering a neutron star/magnetar energizing a young superluminous supernova, or a system with a pulsar/magnetar in the vicinity of a massive black hole are the most plausible ones to date.
More recent observations have shown that the bursts from FRB~121102 are almost 100\% linearly polarized at an unexpectedly high and variable Faraday rotation measure, that has been observed to date only in vicinities of massive/supermassive black holes. The bursts are thus likely produced from a neutron star in such environment, although the system can still be explained by a young neutron star embedded in a highly magnetized pulsar wind nebula or supernova remnant.
New FRB localizations would unveil if this source is representative of the whole population or a particular case.
Upcoming interferometric searches are expected to report tens of these localizations in the coming years, dramatically boosting the field of FRBs.
}

\FullConference{14th European VLBI Network Symposium \& Users Meeting (EVN 2018)\\
		8-11 October 2018\\
		Granada, Spain}

\begin{document}

\section{Introduction}

Fast Radio Bursts (FRBs) are extragalactic transient sources of unknown physical origin \cite{lorimer2007,thornton2013,petroff2016}. These events exhibit millisecond-duration radio flashes that are highly dispersed, with dispersion measures (DMs) consistent with extragalactic origins.
The first FRB was discovered in 2007 after a re-analysis of data from the Parkes Radio Telescope Galactic pulsar survey. A single signal with a peak brightness of $\sim30$ Jy lasting $\sim 5$ ms was detected, and the inferred DM suggested a distance to the source of the order of 1~Gpc \cite{lorimer2007}. This value for the distance would imply an extremely luminous event that could not be understood by any known system.

The field did not boost until more of such events were clearly found \cite{thornton2013}.
Nowadays, tens of FRBs have been reported from several single-dish telescopes \cite{petroff2016,keane2018}. However, the origin of FRBs remains unknown, although a large number of possible scenarios has been proposed so far (see e.g. \cite{katz2016,katz2018,pen2018}).

Although the estimated burst energies are comparable to the spin-down energy of known pulsars such as the Crab, the observed burst luminosities are $\sim 10^{10\text{--}15}$ times larger than the typical ones of individual pulses from pulsars.
Candidate classes of objects that liberate this much (and even much higher) total energies are gamma-ray bursts (GRBs) and supernova (SN) explosions. However, in these cases the derived rate of events ($10^{3\text{--}4}\ \mathrm{sky^{-1}\ d^{-1}}$) \cite{keane2015,law2015} does not match the ones in GRBs or SNe. In addition, multiwavelength searches have not revealed robust afterglow detections to date (although candidate events have been reported \cite{keane2016,mahony2018}, these were not confirmed \cite{giroletti2016}).

The interest on FRBs not only resides on their intriguing nature, but on their direct and indirect implications on several astrophysical fields. If FRBs are detected in great numbers at cosmological distances, they could shed light on the distribution of baryonic matter in the Universe, and the nature of dark energy \cite{macquart2015}.

\subsection{Fast transients with the European VLBI Network (EVN)}

Very long baseline interferometry (VLBI) observations are ideal to precisely localize transient sources on milliarcsecond (mas) scales and to determine their morphological and temporal evolution. The accurate astrometry obtained from these observations is critical for extragalactic sources, where associations require the identification (and distinction) of local sources within the host galaxies. Simultaneously, the reached (mas) resolution allows the study of morphological changes in the transient events (e.g. the evolution of afterglows) on timescales as short as weeks or years.

The European VLBI Network (EVN)\footnote{See the EVN website: \url{http://www.evlbi.org}.} is a network of antennas spread mainly within Europe, Asia, and Africa. The available baselines (up to 10\,000~km) and the observing frequencies (primarily between 1.4 and 22~GHz) imply a resolution of the order of mas. The data, recorded locally at each station, are sent to The Joint Institute for VLBI ERIC (JIVE) in The Netherlands for correlation with the EVN Software Correlator (SFXC) \cite{keimpema2015}.

One of the operating modes of the EVN is the real-time correlation (e-EVN), where data are directly sent in real-time and electronically to JIVE. Runs of 24~h are conducted on average once per month. This sampling makes the e-EVN an ideal tool for monitoring programs and transient studies. Highlights of synchrotron transients science, and a brief description of the years of work put into detecting and localizing fast transients with the e-EVN are described by \cite{paragi2016}.

\section{The only known repeating FRB}

The Arecibo Telescope discovered its first FRB in 2012, FRB~121102 \cite{spitler2014}. Multiple bursts were soon discovered, making this source the first and only repeating FRB known so far\footnote{A second repeater has just been discovered by CHIME \cite{chime2019}.} \cite{spitler2016,scholz2016}. The measured DM of $\sim 560\ \mathrm{pc\ cm^{-3}}$ implied a column density around a factor of three larger than the expected one from the contribution of our Galaxy in that direction (which is close to the Galactic anticenter). This made the source of a likely extragalactic origin.

Despite multiple bursts have been detected, no periodicities have been found so far. However, there are active periods when a higher number of bursts is observed. Most of the bursts have been seen at $\sim 1.4$~GHz, but there have been detections at 4--8~GHz as well \cite{gajjar2018,spitler2018}. The bursts are typically narrow-band, with characteristic widths of $\sim 500\ \mathrm{MHz}$ \cite{law2017}. Complex time/frequency structures and the existence of sub-bursts as narrow as $\sim 30\ \mathrm{\upmu s}$ have also been revealed \cite{hessels2018}. It remains unclear if these structures are intrinsic \cite{hessels2018}, or they appear due to propagation effects in the local medium \cite{cordes2017}.

Although similar structures have been reported in other FRBs, no repetitions have been observed in any other FRB to date (2018), even after exhaustive searches \cite{shannon2018}. This fact has raised questions about FRB~121102 being representative or not of the whole population of FRBs \cite{palaniswamy2018}.

\section{The precise localization of FRB~121102}

The existence of multiple bursts in FRB~121102 allowed the scheduling of interferometric observations on a FRB for the first time. The Arecibo detections provided an a-priori localization with an uncertainty of a few arcminutes, enough to fit within the primary beam of typical dishes.

\begin{figure}
	\includegraphics[width=\textwidth]{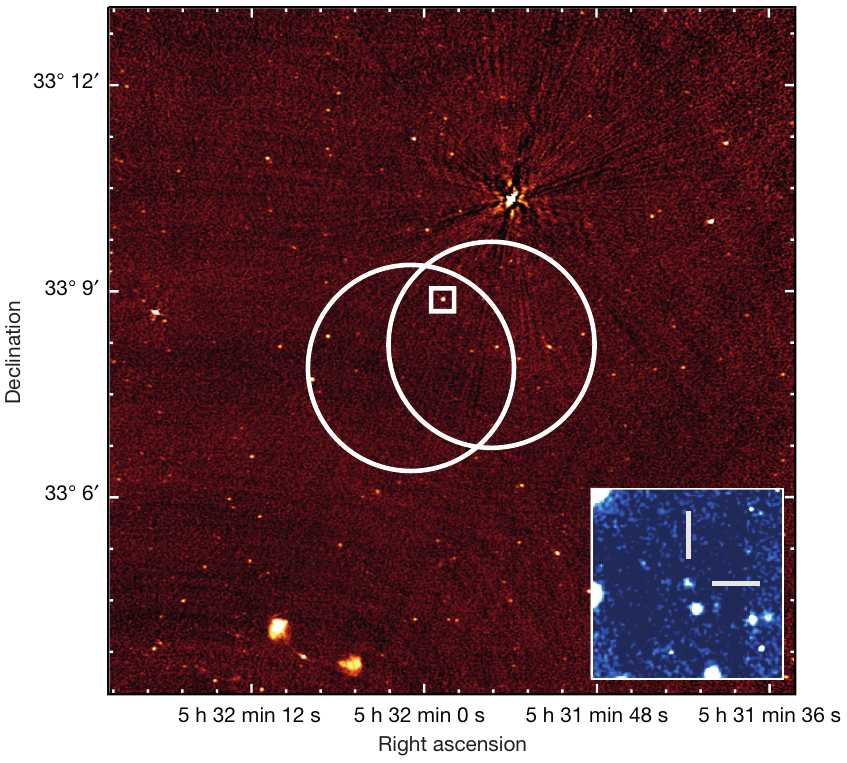}
	\caption{Radio and optical images of the field around FRB~121102. The main figure shows the field as seen by the VLA at 3~GHz, with a resolution of 2~arcsec and a rms noise level of $2\ \mathrm{\upmu Jy\ beam^{-1}}$. The white square is centered at the position where the bursts are located, and the white circles represent previous Arecibo burst detections and their uncertainty regions. The secondary figure represents a zoom on the white square showing the optical image as seen by Gemini at {\em r} band. The two lines point to the position of FRB~121102. Persistent radio and optical counterparts are detected coincident with such position. See \cite{chatterjee2017}.}
	\label{fig:vla}
\end{figure}

Two teams initiated interferometric observations of FRB~121102 independently: an US-based team with the Karl G. Jansky Very Large Array (VLA), starting in November 2015, and a European collaboration using the EVN (including the Arecibo Telescope), in February 2016. Five epochs of e-EVN observations during the spring showed no bursts.
In the summer of that year, high activity was reported from Arecibo single-dish observations. The subsequent VLA (and later EVN) follow-up observations resulted in the detection of several bursts, and their association at the sub-arcsecond accuracy level with a faint (180-$\mathrm{\upmu Jy}$ at 3~GHz) persistent radio source that was compact even on EVN scales \cite{chatterjee2017} (see Figure~\ref{fig:vla}). These bursts appeared $\sim 10^{4\text{--}5}$ times brighter than the persistent emission. There was only low-level ($\sim 10\%$) variability observed in the persistent radio source and it did not correlate with the bursts, therefore there was no sign of relativistic outflows. The spectrum of this source also remains stable
along the time, with a relatively flat emission between 1 and 10~GHz and a cutoff of unclear origin at higher frequencies \cite{chatterjee2017}.

We detected bursts from FRB~121102 in one observation with the EVN in September 2016. Four bursts were detected, with one of them around ten times brighter than the other ones ($\sim 4\ \mathrm{Jy}$ to be compared to $\sim 0.2$--$0.5\ \mathrm{Jy}$). These data allowed us to study both the persistent source and the burst location on milliarcsecond scales \cite{marcote2017}.

\begin{figure}[t]
	\includegraphics[width=\textwidth]{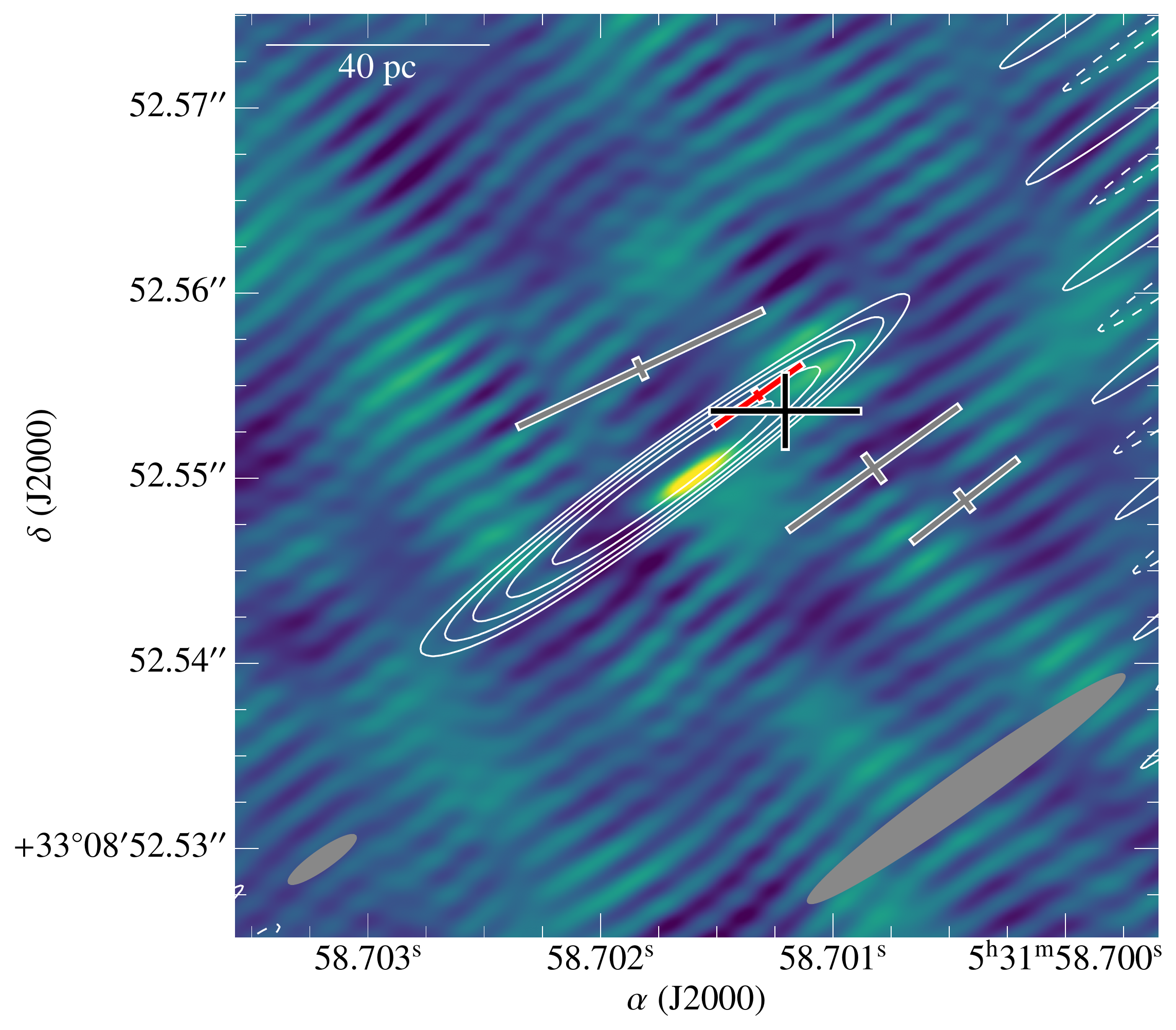}
	\caption{Radio images of the field around FRB~121102 as seen by the EVN at 1.7~GHz with a resolution of $21\times 2\ \mathrm{mas^2}$ (represented by the white contours, that start at the $2$-$\sigma$ noise level of $10\ \mathrm{\upmu Jy\ beam^{-1}}$ and increase by factors of $2^{1/2}$), and at 5.0~GHz with a resolution of $4 \times 1\ \mathrm{mas^2}$ and a rms of $14\ \mathrm{\upmu Jy\ beam^{-1}}$ (represented by the colorscale). The synthesized beams at each frequency are represented by the gray ellipses at the right and left bottom corners, respectively. The crosses represent the positions of the single bursts and their lengths represent the 1-$\sigma$ statistical uncertainty in each direction. The red crosses represent the strongest burst detected by the EVN, which is expected to exhibit the most accurate astrometry, and the gray ones represent the other three bursts detected during the same observation. The black cross represents the weighted average position of the bursts, used for further discussions. See \cite{marcote2017}.}
	\label{fig:evn}
\end{figure}

Whereas the bursts were detected in a 1.7-GHz observation, the persistent source was imaged in both 1.7 and 5.0-GHz observations. The data at the highest frequency allowed us to put strong constraints on the source size ($< 0.7\ \mathrm{pc}$). Simultaneously, we provided accurate localizations for the four detected bursts. Whereas the strongest burst produced the most accurate measurement, we determined a weighted average position from all of them. This average position is coincident with the persistent source within a projected separation of $< 40\ \mathrm{pc}$ at 95\% confidence level \cite{marcote2017} (see Figure~\ref{fig:evn}). The obtained separation constraint suggests a direct physical connection between the bursts and the persistent source.
The robustness of the burst astrometry was tested by a detailed analysis of single-pulse imaging from data of the known pulsar B0525+21, which was observed within these EVN observations \cite{marcote2017}.

The data recorded by Arecibo during the EVN observations also allowed us to study the narrow-band frequency structure of the bursts. A fine-scale frequency structure of $\sim 58\ \mathrm{kHz}$ was measured, with is consistent with Galactic scintillation when the light enters the Galaxy \cite{hessels2018}.

\begin{figure}[t]
	\includegraphics[width=\textwidth]{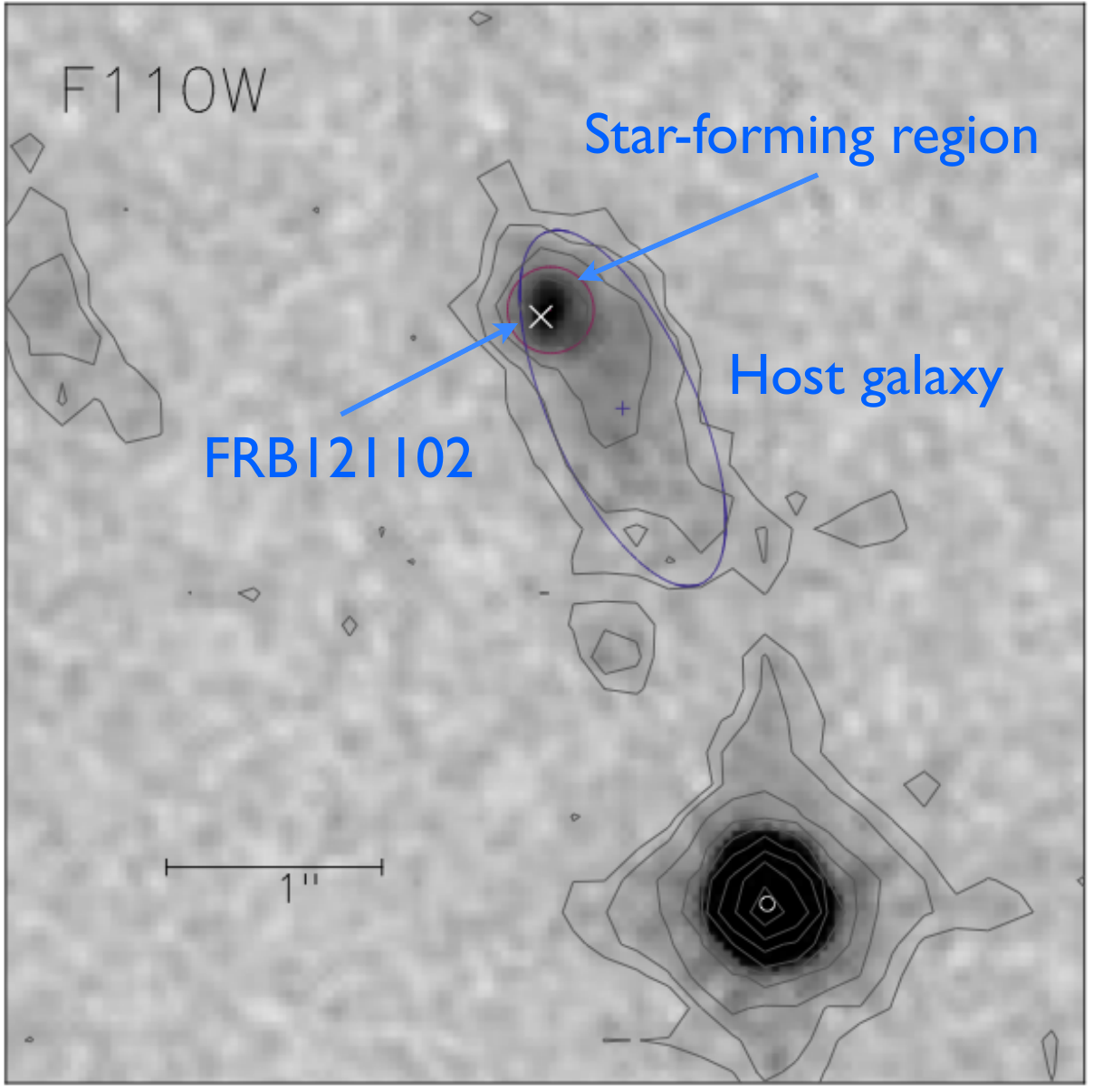}
	\caption{Optical image of the field around FRB~121102 as seen by the {\em HST}/WFC3 at {\em J} band. FRB~121102 (represented by a yellow cross) is coincident with a star-forming region (highlighted by the red circle) that dominates the optical emission of the host, dwarf, galaxy. The purple ellipse highlights the mean extension of the galaxy. The strong object at the right bottom is a field star. Adapted from \cite{bassa2017}.}
	\label{fig:hst}
\end{figure}
Optical observations with the Gemini \cite{tendulkar2017} and {\em HST} \cite{bassa2017} were conducted to unveil the host of FRB~121102. These data unveiled a low-metallicity dwarf ($< 10\ \mathrm{kpc}$ in diameter) galaxy located at a redshift of $\sim 0.193$ (luminosity distance of 972~Mpc). The bursts, and the associated radio persistent source, were located inside a star-forming region with a radius of 1.2~kpc that dominates the optical emission of the whole galaxy (see Figure~\ref{fig:hst}).
The properties of this galaxy resemble the ones where hydrogen-poor superluminous supernovae (SLSN) and long-duration gamma-ray bursts (LGRBs) are typically located. This suggests a possible connection between FRBs and these sources.

At other wavelengths, deep optical \cite{hardy2017,magic2018}, X-ray and GeV gamma-ray \cite{scholz2017}, and TeV gamma-ray \cite{magic2018} burst searches have been conducted, with null results so far. It thus remains unclear if FRBs exhibit pulsed emission outside the radio wavelengths.

Detailed observations of the bursts from FRB~121102 at 5~GHz revealed the presence of an unexpectedly high Faraday rotation measure (RM) of $\sim 1.4 \times 10^5\ \mathrm{rad\ m^{-2}}$, and a 100\% linear polarization in the bursts \cite{michilli2018}. Such large RM values have only been reported so far in the vicinity of Sagittarius~A$^{\ast}$, i.e. in the surroundings of massive black holes.

\section{Scenarios explaining FRB~121102}

Any model expected to explain FRB~121102 needs to account for the existence of bursts and the presence of the compact and persistent radio source. Most probable scenarios assume a young (few decades old) and energetic magnetar (or highly magnetized neutron star) origin \cite{connor2016,cordes2016}. These scenarios would naturally explain the observed bursts, which could likely be originated by coherent curvature radiation \cite{ghisellini2017}.

The persistent source would then be explained by the presence of a superluminous supernova (SLSN) \cite{murase2016,piro2016,kashiyama2017,margalit2018}. Although this would explain the co-localization of the bursts within the persistent source, and matches the environment where FRB~121102 is located (low-metallicity star-forming dwarf galaxy), the properties of the persistent source still challenge this explanation (such as its luminosity or lack of long-term variability).

Other scenarios consider the presence of a young magnetar which is interacting with the jet or surroundings of a massive black hole \cite{pen2015,cordes2016,zhang2017,zhang2018}. These scenarios would naturally explain the existence of bursts (from the magnetar), the properties of the persistent source, and the high RM. Some theoretical models considering that bursts are originated due to strong plasma turbulence within the jet of a massive black hole \cite{vieyro2017} or as synchrotron masers \cite{ghisellini2017} have also been proposed, but these ones present a reduced observational support. In addition, there are a large number of speculative scenarios \cite{pen2018}.

\section{Conclusions and future prospects}

The FRB field has progressed a lot in the past few years. Several tens of these objects have been found, but the existence of only one repeater\footnote{A second repeater has just been discovered by CHIME, with no clear counterparts found yet \cite{chime2019}.} and only one precise localization hampers our understanding of these sources.
It remains unclear if repeating and non-repeating FRBs belong to two different types of systems, or if the repetition of bursts is only subject to particular conditions in similar systems.

The available data do not support the presence of afterglows after the creation of bursts, having strong implications on future FRB searches. New FRB localizations, which are fundamental to unveil the nature of these objects, would require interferometric observations at the arrival burst times. Only detections of bursts during high resolution (interferometric) observations would allow us to associate the bursts to persistent sources, as opposed to other fields such as gravitational wave events (where the presence of afterglows allows us to locate the event from later observations).

The precise localization of FRB~121102 has led to the first detailed studies of a FRB. We have unveiled the environment of this source to exhibit extreme conditions and to be associated with a persistent and compact radio source embedded in a low-metallicity star-forming region of a dwarf galaxy. Genuine localizations of more FRBs are mandatory to unveil their hosts and trace their properties that are common to the whole population.

Upcoming facilities such as WSRT/Apertif, ASKAP, CHIME, or UTMOST have dedicated searches for new FRBs. These searches will lead to new associations that will allow us to understand the uniqueness of FRB~121102 among the whole population. However, we note that the resolution reached by VLBI observations would be in any case mandatory to study in detail the local environment to the FRBs in their host galaxies. The same methodology we have developed for the detection and localization of fast transients in general \cite{paragi2016}, and the repeater FRB~121102 in particular \cite{marcote2017}, could be used for future localizations of repeating FRBs.

Direct VLBI imaging of single pulses is also a (remote) possibility. However, blind commensal searches for FRBs in single dish data obtained during Very Long Baseline Array (VLBA) experiments have not been successful to date yet \cite{burkespolaor2016}. While the EVN has much more sensitive individual dishes to look for single pulses, these have smaller field of views as well. Therefore it is not clear how productive single-pulse FRB searches could be with the EVN. From the current FRB rates we estimate that if the EVN Archive contained full field of view raw data for all experiments since its existence ($\sim 1999$--$2000$), we could have identified and localized a few more FRBs at least from archival EVN observations.

\bibliographystyle{JHEP}
\bibliography{/Users/hawky/Documents/Reference/bibliography.bib}

\providecommand{\href}[2]{#2}\begingroup\raggedright\begin{thebibliography}{10}

\bibitem{lorimer2007}
D.~R. {Lorimer}, M.~{Bailes}, M.~A. {McLaughlin}, D.~J. {Narkevic} and
  F.~{Crawford}, \emph{{A Bright Millisecond Radio Burst of Extragalactic
  Origin}}, \href{https://doi.org/10.1126/science.1147532}{\emph{Science}
  {\bfseries 318} (2007) 777}
  [\href{https://arxiv.org/abs/0709.4301}{{\ttfamily 0709.4301}}].

\bibitem{thornton2013}
D.~{Thornton}, B.~{Stappers}, M.~{Bailes}, B.~{Barsdell}, S.~{Bates}, N.~D.~R.
  {Bhat} et~al., \emph{{A Population of Fast Radio Bursts at Cosmological
  Distances}}, \href{https://doi.org/10.1126/science.1236789}{\emph{Science}
  {\bfseries 341} (2013) 53} [\href{https://arxiv.org/abs/1307.1628}{{\ttfamily
  1307.1628}}].

\bibitem{petroff2016}
E.~{Petroff}, E.~D. {Barr}, A.~{Jameson}, E.~F. {Keane}, M.~{Bailes},
  M.~{Kramer} et~al., \emph{{FRBCAT: The Fast Radio Burst Catalogue}},
  \href{https://doi.org/10.1017/pasa.2016.35}{\emph{\pasa} {\bfseries 33}
  (2016) e045} [\href{https://arxiv.org/abs/1601.03547}{{\ttfamily
  1601.03547}}].

\bibitem{keane2018}
E.~F. {Keane}, \emph{{The future of fast radio burst science}},
  \href{https://doi.org/10.1038/s41550-018-0603-0}{\emph{Nature Astronomy}
  {\bfseries 2} (2018) 865} [\href{https://arxiv.org/abs/1811.00899}{{\ttfamily
  1811.00899}}].

\bibitem{katz2016}
J.~I. {Katz}, \emph{{Fast radio bursts --- A brief review: Some questions,
  fewer answers}},
  \href{https://doi.org/10.1142/S0217732316300135}{\emph{Modern Physics Letters
  A} {\bfseries 31} (2016) 1630013}
  [\href{https://arxiv.org/abs/1604.01799}{{\ttfamily 1604.01799}}].

\bibitem{katz2018}
J.~I. {Katz}, \emph{{Fast radio bursts}},
  \href{https://doi.org/10.1016/j.ppnp.2018.07.001}{\emph{Progress in Particle
  and Nuclear Physics} {\bfseries 103} (2018) 1}
  [\href{https://arxiv.org/abs/1804.09092}{{\ttfamily 1804.09092}}].

\bibitem{pen2018}
U.-L. {Pen}, \emph{{The nature of fast radio bursts}},
  \href{https://doi.org/10.1038/s41550-018-0620-z}{\emph{Nature Astronomy}
  {\bfseries 2} (2018) 842} [\href{https://arxiv.org/abs/1811.00605}{{\ttfamily
  1811.00605}}].

\bibitem{keane2015}
E.~F. {Keane} and E.~{Petroff}, \emph{{Fast radio bursts: search sensitivities
  and completeness}},
  \href{https://doi.org/10.1093/mnras/stu2650}{\emph{\mnras} {\bfseries 447}
  (2015) 2852} [\href{https://arxiv.org/abs/1409.6125}{{\ttfamily 1409.6125}}].

\bibitem{law2015}
C.~J. {Law}, G.~C. {Bower}, S.~{Burke-Spolaor}, B.~{Butler}, E.~{Lawrence},
  T.~J.~W. {Lazio} et~al., \emph{{A Millisecond Interferometric Search for Fast
  Radio Bursts with the Very Large Array}},
  \href{https://doi.org/10.1088/0004-637X/807/1/16}{\emph{\apj} {\bfseries 807}
  (2015) 16} [\href{https://arxiv.org/abs/1412.7536}{{\ttfamily 1412.7536}}].

\bibitem{keane2016}
E.~F. {Keane}, S.~{Johnston}, S.~{Bhandari}, E.~{Barr}, N.~D.~R. {Bhat},
  M.~{Burgay} et~al., \emph{{The host galaxy of a fast radio burst}},
  \href{https://doi.org/10.1038/nature17140}{\emph{\nat} {\bfseries 530} (2016)
  453} [\href{https://arxiv.org/abs/1602.07477}{{\ttfamily 1602.07477}}].

\bibitem{mahony2018}
E.~K. {Mahony}, R.~D. {Ekers}, J.-P. {Macquart}, E.~M. {Sadler}, K.~W.
  {Bannister}, S.~{Bhandari} et~al., \emph{{A Search for the Host Galaxy of FRB
  171020}}, \href{https://doi.org/10.3847/2041-8213/aae7cb}{\emph{\apj}
  {\bfseries 867} (2018) L10}
  [\href{https://arxiv.org/abs/1810.04354}{{\ttfamily 1810.04354}}].

\bibitem{giroletti2016}
M.~{Giroletti}, B.~{Marcote}, M.~A. {Garrett}, Z.~{Paragi}, J.~{Yang},
  K.~{Hada} et~al., \emph{{FRB 150418: clues to its nature from European VLBI
  Network and e-MERLIN observations}},
  \href{https://doi.org/10.1051/0004-6361/201629172}{\emph{\aap} {\bfseries
  593} (2016) L16} [\href{https://arxiv.org/abs/1609.01419}{{\ttfamily
  1609.01419}}].

\bibitem{macquart2015}
J.~P. {Macquart}, E.~{Keane}, K.~{Grainge}, M.~{McQuinn}, R.~{Fender},
  J.~{Hessels} et~al., \emph{{Fast Transients at Cosmological Distances with
  the SKA}}, {\emph{Advancing Astrophysics with the Square Kilometre Array
  (AASKA14)} (2015) 55} [\href{https://arxiv.org/abs/1501.07535}{{\ttfamily
  1501.07535}}].

\bibitem{keimpema2015}
A.~{Keimpema}, M.~M. {Kettenis}, S.~V. {Pogrebenko}, R.~M. {Campbell},
  G.~{Cim{\'o}}, D.~A. {Duev} et~al., \emph{{The SFXC software correlator for
  very long baseline interferometry: algorithms and implementation}},
  \href{https://doi.org/10.1007/s10686-015-9446-1}{\emph{Experimental
  Astronomy} {\bfseries 39} (2015) 259}.

\bibitem{paragi2016}
Z.~{Paragi}, \emph{{Transient science with the e-EVN}}, {\emph{ArXiv e-prints}
  (2016) } [\href{https://arxiv.org/abs/1612.00508}{{\ttfamily 1612.00508}}].

\bibitem{spitler2014}
L.~G. {Spitler}, J.~M. {Cordes}, J.~W.~T. {Hessels}, D.~R. {Lorimer}, M.~A.
  {McLaughlin}, S.~{Chatterjee} et~al., \emph{{Fast Radio Burst Discovered in
  the Arecibo Pulsar ALFA Survey}},
  \href{https://doi.org/10.1088/0004-637X/790/2/101}{\emph{\apj} {\bfseries
  790} (2014) 101} [\href{https://arxiv.org/abs/1404.2934}{{\ttfamily
  1404.2934}}].

\bibitem{chime2019}
M.~{Amiri}, K.~{Bandura}, M.~{Bhardwaj}, P.~{Boubel}, M.~M. {Boyce}, P.~J.
  {Boyle} et~al., \emph{{A second source of repeating fast radio bursts}},
  \href{https://doi.org/10.1038/s41586-018-0864-x}{\emph{\nat} (2019)}.

\bibitem{spitler2016}
L.~G. {Spitler}, P.~{Scholz}, J.~W.~T. {Hessels}, S.~{Bogdanov}, A.~{Brazier},
  F.~{Camilo} et~al., \emph{{A repeating fast radio burst}},
  \href{https://doi.org/10.1038/nature17168}{\emph{\nat} {\bfseries 531} (2016)
  202} [\href{https://arxiv.org/abs/1603.00581}{{\ttfamily 1603.00581}}].

\bibitem{scholz2016}
P.~{Scholz}, L.~G. {Spitler}, J.~W.~T. {Hessels}, S.~{Chatterjee}, J.~M.
  {Cordes}, V.~M. {Kaspi} et~al., \emph{{The Repeating Fast Radio Burst FRB
  121102: Multi-wavelength Observations and Additional Bursts}},
  \href{https://doi.org/10.3847/1538-4357/833/2/177}{\emph{\apj} {\bfseries
  833} (2016) 177} [\href{https://arxiv.org/abs/1603.08880}{{\ttfamily
  1603.08880}}].

\bibitem{gajjar2018}
V.~{Gajjar}, A.~P.~V. {Siemion}, D.~C. {Price}, C.~J. {Law}, D.~{Michilli},
  J.~W.~T. {Hessels} et~al., \emph{{Highest Frequency Detection of FRB 121102
  at 4-8 GHz Using the Breakthrough Listen Digital Backend at the Green Bank
  Telescope}}, \href{https://doi.org/10.3847/1538-4357/aad005}{\emph{\apj}
  {\bfseries 863} (2018) 2} [\href{https://arxiv.org/abs/1804.04101}{{\ttfamily
  1804.04101}}].

\bibitem{spitler2018}
L.~G. {Spitler}, W.~{Herrmann}, G.~C. {Bower}, S.~{Chatterjee}, J.~M. {Cordes},
  J.~W.~T. {Hessels} et~al., \emph{{Detection of Bursts from FRB~121102~with
  the Effelsberg 100 m Radio Telescope at 5 GHz and the Role of
  Scintillation}}, \href{https://doi.org/10.3847/1538-4357/aad332}{\emph{\apj}
  {\bfseries 863} (2018) 150}
  [\href{https://arxiv.org/abs/1807.03722}{{\ttfamily 1807.03722}}].

\bibitem{law2017}
C.~J. {Law}, M.~W. {Abruzzo}, C.~G. {Bassa}, G.~C. {Bower}, S.~{Burke-Spolaor},
  B.~J. {Butler} et~al., \emph{{A Multi-telescope Campaign on FRB 121102:
  Implications for the FRB Population}}, {\emph{ArXiv e-prints} (2017) }
  [\href{https://arxiv.org/abs/1705.07553}{{\ttfamily 1705.07553}}].

\bibitem{hessels2018}
J.~W.~T. {Hessels}, L.~G. {Spitler}, A.~D. {Seymour}, J.~M. {Cordes},
  D.~{Michilli}, R.~S. {Lynch} et~al., \emph{{FRB 121102 Bursts Show Complex
  Time-Frequency Structure}}, {\emph{arXiv e-prints} (2018) arXiv:1811.10748}
  [\href{https://arxiv.org/abs/1811.10748}{{\ttfamily 1811.10748}}].

\bibitem{cordes2017}
J.~M. {Cordes}, I.~{Wasserman}, J.~W.~T. {Hessels}, T.~J.~W. {Lazio},
  S.~{Chatterjee} and R.~S. {Wharton}, \emph{{Lensing of Fast Radio Bursts by
  Plasma Structures in Host Galaxies}},
  \href{https://doi.org/10.3847/1538-4357/aa74da}{\emph{\apj} {\bfseries 842}
  (2017) 35} [\href{https://arxiv.org/abs/1703.06580}{{\ttfamily 1703.06580}}].

\bibitem{shannon2018}
R.~M. {Shannon}, J.~P. {Macquart}, K.~W. {Bannister}, R.~D. {Ekers}, C.~W.
  {James}, S.~{Os{\l}owski} et~al., \emph{{The dispersion-brightness relation
  for fast radio bursts from a wide-field survey}},
  \href{https://doi.org/10.1038/s41586-018-0588-y}{\emph{\nat} {\bfseries 562}
  (2018) 386}.

\bibitem{palaniswamy2018}
D.~{Palaniswamy}, Y.~{Li} and B.~{Zhang}, \emph{{Are There Multiple Populations
  of Fast Radio Bursts?}},
  \href{https://doi.org/10.3847/2041-8213/aaaa63}{\emph{\apj} {\bfseries 854}
  (2018) L12} [\href{https://arxiv.org/abs/1703.09232}{{\ttfamily
  1703.09232}}].

\bibitem{chatterjee2017}
S.~{Chatterjee}, C.~J. {Law}, R.~S. {Wharton}, S.~{Burke-Spolaor}, J.~W.~T.
  {Hessels}, G.~C. {Bower} et~al., \emph{{A direct localization of a fast radio
  burst and its host}}, \href{https://doi.org/10.1038/nature20797}{\emph{\nat}
  {\bfseries 541} (2017) 58}
  [\href{https://arxiv.org/abs/1701.01098}{{\ttfamily 1701.01098}}].

\bibitem{marcote2017}
B.~{Marcote}, Z.~{Paragi}, J.~W.~T. {Hessels}, A.~{Keimpema}, H.~J. {van
  Langevelde}, Y.~{Huang} et~al., \emph{{The Repeating Fast Radio Burst FRB
  121102 as Seen on Milliarcsecond Angular Scales}},
  \href{https://doi.org/10.3847/2041-8213/834/2/L8}{\emph{\apjl} {\bfseries
  834} (2017) L8} [\href{https://arxiv.org/abs/1701.01099}{{\ttfamily
  1701.01099}}].

\bibitem{bassa2017}
C.~G. {Bassa}, S.~P. {Tendulkar}, E.~A.~K. {Adams}, N.~{Maddox}, S.~{Bogdanov},
  G.~C. {Bower} et~al., \emph{{FRB 121102 Is Coincident with a Star-forming
  Region in Its Host Galaxy}},
  \href{https://doi.org/10.3847/2041-8213/aa7a0c}{\emph{\apjl} {\bfseries 843}
  (2017) L8} [\href{https://arxiv.org/abs/1705.07698}{{\ttfamily 1705.07698}}].

\bibitem{tendulkar2017}
S.~P. {Tendulkar}, C.~G. {Bassa}, J.~M. {Cordes}, G.~C. {Bower}, C.~J. {Law},
  S.~{Chatterjee} et~al., \emph{{The Host Galaxy and Redshift of the Repeating
  Fast Radio Burst FRB 121102}},
  \href{https://doi.org/10.3847/2041-8213/834/2/L7}{\emph{\apjl} {\bfseries
  834} (2017) L7} [\href{https://arxiv.org/abs/1701.01100}{{\ttfamily
  1701.01100}}].

\bibitem{hardy2017}
L.~K. {Hardy}, V.~S. {Dhillon}, L.~G. {Spitler}, S.~P. {Littlefair}, R.~P.
  {Ashley}, A.~{De Cia} et~al., \emph{{A search for optical bursts from the
  repeating fast radio burst FRB 121102}},
  \href{https://doi.org/10.1093/mnras/stx2153}{\emph{\mnras} {\bfseries 472}
  (2017) 2800} [\href{https://arxiv.org/abs/1708.06156}{{\ttfamily
  1708.06156}}].

\bibitem{magic2018}
{MAGIC Collaboration}, V.~A. {Acciari}, S.~{Ansoldi}, L.~A. {Antonelli},
  A.~{Arbet Engels}, C.~{Arcaro} et~al., \emph{{Constraining very-high-energy
  and optical emission from FRB 121102 with the MAGIC telescopes}},
  \href{https://doi.org/10.1093/mnras/sty2422}{\emph{\mnras} {\bfseries 481}
  (2018) 2479} [\href{https://arxiv.org/abs/1809.00663}{{\ttfamily
  1809.00663}}].

\bibitem{scholz2017}
P.~{Scholz}, S.~{Bogdanov}, J.~W.~T. {Hessels}, R.~S. {Lynch}, L.~G. {Spitler},
  C.~G. {Bassa} et~al., \emph{{Simultaneous X-Ray, Gamma-Ray, and Radio
  Observations of the Repeating Fast Radio Burst FRB 121102}},
  \href{https://doi.org/10.3847/1538-4357/aa8456}{\emph{\apj} {\bfseries 846}
  (2017) 80} [\href{https://arxiv.org/abs/1705.07824}{{\ttfamily 1705.07824}}].

\bibitem{michilli2018}
D.~{Michilli}, A.~{Seymour}, J.~W.~T. {Hessels}, L.~G. {Spitler}, V.~{Gajjar},
  A.~M. {Archibald} et~al., \emph{{An extreme magneto-ionic environment
  associated with the fast radio burst source FRB 121102}},
  \href{https://doi.org/10.1038/nature25149}{\emph{\nat} {\bfseries 553} (2018)
  182}.

\bibitem{connor2016}
L.~{Connor}, J.~{Sievers} and U.-L. {Pen}, \emph{{Non-cosmological FRBs from
  young supernova remnant pulsars}},
  \href{https://doi.org/10.1093/mnrasl/slv124}{\emph{\mnras} {\bfseries 458}
  (2016) L19} [\href{https://arxiv.org/abs/1505.05535}{{\ttfamily
  1505.05535}}].

\bibitem{cordes2016}
J.~M. {Cordes} and I.~{Wasserman}, \emph{{Supergiant pulses from extragalactic
  neutron stars}}, \href{https://doi.org/10.1093/mnras/stv2948}{\emph{\mnras}
  {\bfseries 457} (2016) 232}
  [\href{https://arxiv.org/abs/1501.00753}{{\ttfamily 1501.00753}}].

\bibitem{ghisellini2017}
G.~{Ghisellini} and N.~{Locatelli}, \emph{{Coherent curvature radiation and
  Fast Radio Bursts}}, {\emph{ArXiv e-prints} (2017) }
  [\href{https://arxiv.org/abs/1708.07507}{{\ttfamily 1708.07507}}].

\bibitem{murase2016}
K.~{Murase}, K.~{Kashiyama} and P.~{M{\'e}sz{\'a}ros}, \emph{{A burst in a wind
  bubble and the impact on baryonic ejecta: high-energy gamma-ray flashes and
  afterglows from fast radio bursts and pulsar-driven supernova remnants}},
  \href{https://doi.org/10.1093/mnras/stw1328}{\emph{\mnras} {\bfseries 461}
  (2016) 1498} [\href{https://arxiv.org/abs/1603.08875}{{\ttfamily
  1603.08875}}].

\bibitem{piro2016}
A.~L. {Piro}, \emph{{The Impact of a Supernova Remnant on Fast Radio Bursts}},
  \href{https://doi.org/10.3847/2041-8205/824/2/L32}{\emph{\apjl} {\bfseries
  824} (2016) L32} [\href{https://arxiv.org/abs/1604.04909}{{\ttfamily
  1604.04909}}].

\bibitem{kashiyama2017}
K.~{Kashiyama} and K.~{Murase}, \emph{{Testing the Young Neutron Star Scenario
  with Persistent Radio Emission Associated with FRB 121102}},
  \href{https://doi.org/10.3847/2041-8213/aa68e1}{\emph{\apjl} {\bfseries 839}
  (2017) L3} [\href{https://arxiv.org/abs/1701.04815}{{\ttfamily 1701.04815}}].

\bibitem{margalit2018}
B.~{Margalit}, B.~D. {Metzger}, E.~{Berger}, M.~{Nicholl}, T.~{Eftekhari} and
  R.~{Margutti}, \emph{{Unveiling the engines of fast radio bursts,
  superluminous supernovae, and gamma-ray bursts}},
  \href{https://doi.org/10.1093/mnras/sty2417}{\emph{\mnras} {\bfseries 481}
  (2018) 2407} [\href{https://arxiv.org/abs/1806.05690}{{\ttfamily
  1806.05690}}].

\bibitem{pen2015}
U.-L. {Pen} and L.~{Connor}, \emph{{Local Circumnuclear Magnetar Solution to
  Extragalactic Fast Radio Bursts}},
  \href{https://doi.org/10.1088/0004-637X/807/2/179}{\emph{\apj} {\bfseries
  807} (2015) 179} [\href{https://arxiv.org/abs/1501.01341}{{\ttfamily
  1501.01341}}].

\bibitem{zhang2017}
B.-B. {Zhang} and B.~{Zhang}, \emph{{Repeating FRB 121102: Eight-year Fermi-LAT
  Upper Limits and Implications}},
  \href{https://doi.org/10.3847/2041-8213/aa7633}{\emph{\apjl} {\bfseries 843}
  (2017) L13} [\href{https://arxiv.org/abs/1705.04242}{{\ttfamily
  1705.04242}}].

\bibitem{zhang2018}
B.~{Zhang}, \emph{{FRB 121102: A Repeatedly Combed Neutron Star by a Nearby
  Low-luminosity Accreting Supermassive Black Hole}},
  \href{https://doi.org/10.3847/2041-8213/aaadba}{\emph{\apj} {\bfseries 854}
  (2018) L21} [\href{https://arxiv.org/abs/1801.05436}{{\ttfamily
  1801.05436}}].

\bibitem{vieyro2017}
F.~L. {Vieyro}, G.~E. {Romero}, V.~{Bosch-Ramon}, B.~{Marcote} and M.~V. {del
  Valle}, \emph{{A model for the repeating FRB 121102 in the AGN scenario}},
  \href{https://doi.org/10.1051/0004-6361/201730556}{\emph{\aap} {\bfseries
  602} (2017) A64} [\href{https://arxiv.org/abs/1704.08097}{{\ttfamily
  1704.08097}}].

\bibitem{burkespolaor2016}
S.~{Burke-Spolaor}, C.~M. {Trott}, W.~F. {Brisken}, A.~T. {Deller}, W.~A.
  {Majid}, D.~{Palaniswamy} et~al., \emph{{Limits on Fast Radio Bursts from
  Four Years of the V-FASTR Experiment}},
  \href{https://doi.org/10.3847/0004-637X/826/2/223}{\emph{\apj} {\bfseries
  826} (2016) 223} [\href{https://arxiv.org/abs/1605.07606}{{\ttfamily
  1605.07606}}].

\end{thebibliography}\endgroup

%
%

\end{document}